\documentclass[aps, pra, preprintnumbers,notitlepage, superscriptaddress, twocolumn]{revtex4-1}
\usepackage{amsmath, amsthm, amssymb}
\usepackage{pifont}
\usepackage{dsfont}
\usepackage{amscd}
\usepackage{amsfonts}
\usepackage{enumitem}
\usepackage{textcomp}
\usepackage{float}
\usepackage{mathpazo}
\usepackage{hyperref}
\hypersetup{
    colorlinks,
    citecolor=magenta,
    filecolor=black,
    linkcolor=blue,
    urlcolor=black
}
\fontfamily{palatino-ttf}

\usepackage[pdftex]{graphicx}
\usepackage[usenames,dvipsnames]{color}

\usepackage{mathpazo}
\newcommand{\be}{\begin{equation}}
\newcommand{\ee}{\end{equation}}
\newcommand{\ben}{\begin{enumerate}}
\newcommand{\een}{\end{enumerate}}
\def \bea{\begin{align}}
\def \eea{\end{align}}

\newcommand{\half}{\frac{1}{2}}

\newcommand{\chn}{{\cal N}}

\newcommand{\wt}{\widetilde}

\newcommand{\ra}{\rangle}
\newcommand{\la}{\langle}

\newcommand{\pc}{{\cal P}}

\makeatletter

\newcommand{\Rmnum}[1]{\expandafter\@slowromancap\romannumeral #1@}
\makeatother

\newtheorem{thm}{Theorem}
\newtheorem{lemma}{Lemma}

\newtheorem{cor}[thm]{Corollary}
\newtheorem{defn}[thm]{Definition}

\newtheorem{obs}[thm]{Observation}
\fontfamily{palatino-ttf}

\def\li{\left}
\def\pr{\right}

\providecommand{\er}[1]{{\cal N}_{#1}}

\def\ra{\rangle}
\def\la{\langle}
\def\>{\rangle}
\def\<{\langle}

\def\am{{\cal A}_{M_A}}
\def\bm{{\cal B}_{M_B}}
\def\cm{{\cal C}_{M_C}}

\def\squareforqed{\hbox{\rlap{$\sqcap$}$\sqcup$}}
\def\qed{\ifmmode\squareforqed\else{\unskip\nobreak\hfil
\penalty50\hskip1em\null\nobreak\hfil\squareforqed
\parfillskip=0pt\finalhyphendemerits=0\endgraf}\fi}

\providecommand{\norm}[1]{\lVert#1\rVert}

\begin{document}

\title{Parrondo's paradox and superactivation of classical and quantum capacity of communication channels with memory}

\author{Sergii Strelchuk}
 \email{ss870@cam.ac.uk}
\affiliation{Department of Applied Mathematics and Theoretical Physics, University of Cambridge, Cambridge CB3 0WA, U.K.}

\begin{abstract}
There exist memoryless zero-capacity quantum channels that when used jointly result in the channel with positive capacity. This phenomenon is called superactivation. Making use of Parrondo's paradox, we exhibit examples of superactivation-like effect for the capacity of classical communication channels as well as quantum and private capacity of quantum channels with memory. There are several ingredients necessary for superactivation of quantum capacity to occur in memoryless case. The first one is the requirement for the quantum channels which are amenable for superactivation to come from two distinct families  -- binding entanglement channels and erasure channels. The second one is the ability to utilize inputs which are entangled across the uses of the channels. Our construction uses a single family of erasure channels with classical memory to achieve the same superactivation-like effect for quantum capacity without any of the ingredients above. 

\end{abstract}
\maketitle

When can we achieve reliable communication over an imperfect channel? Classical information theory, developed by Shannon, gives a satisfactory answer to this question when the information that we want to transmit and the underlying channel are both classical. It characterizes the performance of every classical channel with a single figure of merit -- classical capacity. However, if we turn to sending quantum information, it is inadequate because the capacity of quantum channels also depends on what other resources are available. This gives rise to the numerous types of capacity of the quantum channels. In particular, quantum capacity of the quantum channel characterizes its ability to transmit entanglement reliably. Private capacity indicates how suitable is the channel for the task of secret key sharing. The contrast between classical and quantum channels is especially stark when we turn to zero-capacity channels. In classical information theory they are precisely the set of useless channels -- the receiver cannot decode the message reliably from the sender because of the adverse effects of noise. 

In 2008 Smith and Yard~\cite{smith_quantum_2008} showed that this is not the case for quantum memoryless channels. They provided an intriguing example of two zero-capacity memoryless quantum channels, which, when combined, result in reliable transmission of quantum information reversing the deleterious effects of noise. This phenomenon is called {\it superactivation}. It is of great interest because it may provide clues about novel methods of transmitting the frail quantum data in the presence of the strongly deleterious noise in the channel.

One important aspect for the channels whose capacity is currently known to be superactivated is the requirement that they must come from two distinct classes: the first one comes from the set of binding entanglement channels~\cite{horodecki_binding_2000}, and the second one comes either from the set of zero-capacity erasure channels with the rational erasure probability or depolarizing channels in the regime when they become antidegradable~\cite{brandao_when_2012}. The capacity of the channel from each of those classes cannot be superactivated by another channel within same class~\cite{strelchuk_hybrid_2012}. Apart from these two families of channels, it is not known if there exist qualitatively different zero-capacity channels that give rise to the superactivation of the quantum capacity. This is in part because proving that a given channel has zero capacity is hard. 

The simplest example of the channel from the latter class is the 50\% erasure channel: half of the time it forwards the input state to the receiver and half of the time it erases the input state sending Bob the erasure flag. The quantum and private capacity of erasure channels is well studied in the regime without feedback. However, despite its simplicity, the quantum capacity under backward classical communication assistance is unknown~\cite{bennett_capacities_1997, leung_capacity_2009}.

Another crucial ingredient for superactivation which was implicitly present in all of the protocols is the use of inputs, which are entangled across multiple uses of the channels~\cite{smith_quantum_2008}. This manifestly quantum feature was known to increase quantum and classical capacity of quantum channels. It is also the reason why there is no computationally tractable way of finding the capacity of general quantum channel. The extent to which inputs entangled across the uses of the channel may help is unknown.

Unlike the situation with the quantum capacity, showing that the private capacity can be superactivated turned out to be a much harder question that remained without an answer both for the channels with and without memory, despite many efforts~\cite{smith_extensive_2009, li_private_2009}. Private capacity was recently found to be non-additive -- the sum of the private capacities of individual channels turned out to be smaller than the private capacity of the joint channel comprised of them~\cite{smith_extensive_2009, li_private_2009}. Superactivation is the strongest form of non-additivity: it occurs when all the individual channel capacities are zero, but the joint capacity is strictly positive. 

Curiously, for all previously known constructions, which exhibit non-additivity of the private capacity it is also necessary to make use of the inputs, which are entangled across the use of the channel. 

For many tasks of information processing considering merely memoryless channels -- quantum or classical -- is not sufficient. For most practical purposes one would like to know how well the channel performs given finitely many of its uses. In this regime the assumption that noise is uncorrelated between channel uses is no longer justified -- the environment may have memory. This led to explosive research activity in classical and quantum information theory that quantifies the capacity of channels with memory. 

In this paper, we provide a general construction for superactivation of the channels with memory. This construction is in the spirit of Parrondo's paradox~\cite{harmer_game_1999}, originally discovered in the context of Brownian ratchets. It occurs when a player has access to two specific coin-flipping betting games, each of which being losing for him. However, contrary to the intuitive expectation, playing them in an alternating fashion results in a winning game.

We show that the presence of classical memory leads to superactivation of capacity of classical and quantum channels. This indicates that the performance of both classical and quantum channels with memory can no longer be characterized with a single figure of merit -- its capacity.
For quantum channels, we demonstrate superactivation without the above requirements which were necessary for the memoryless channels. We introduce two simple zero-capacity quantum channels which make use only of erasure channels: the first one being memoryless erasure channel and the second one being the combination of erasure channels with classical memory. Their convex combination results in another channel with shared classical memory (i.e. in which both channels in the mixture have access to a shared memory register) which has positive quantum capacity. 

Moreover, this effect also holds for the private capacity of the channels with memory. This is especially interesting because due to its simple construction it may provide clues to constructing the long-sought example of superactivation of the private capacity in the memoryless setting, which despite many efforts still remains elusive. The main obstacle for this lies in the difficulty of proving that a given channel has no private capacity.

For classical channels, we slightly modify the construction used to demonstrate superactivation of the quantum and private capacity of the quantum channel. 

Quantum channels with classical memory can be much more powerful when it comes to information transmission. There have been demonstrations of how  memory may benefit  information transmission~\cite{macchiavello_entanglement-enhanced_2002}. 
In practice, as our construction will show, supplementing quantum channels with classical memory offers superior error-correcting capabilities, at a very small price. However, in theory, such channels are hard to analyze: estimating their properties and computing the capacity turns out to be a much harder task than for their memoryless counterparts. In the quantum case this is especially difficult because for quantum channels with memory there exists no notion of state-channel duality. Consequently, one cannot introduce a Choi-Jamio\l kowski state to determine the properties of memoryless channels in the same way as in~\cite{horodecki_binding_2000}, so new techniques are required.

The paper is structured as follows: in Section~\ref{genconstr} we introduce the general construction, which gives rise to the Parrondo effect in the context of communication channels with memory. In Section~\ref{quantcap} we find channels, which demonstrate the superactivation effect for the quantum and private capacity of quantum channels with memory. Lastly, in Section~\ref{classcap}, followed by Discussion, we provide the evidence for the superactivation-like effect for classical channels with memory.

\section{General Construction}\label{genconstr}
We now consider three channels $\am$, $\bm$ and $\cm$, where the subscript $M_{i}\in\mathbb{Z},i=A,B,C$ denotes the memory register of each channel respectively. At this point, these channels may be classical or quantum, and they encapsulate the channels which give rise to Parrondo's paradox. They will differ depending on the type of capacity for which we  demonstrate superactivation and also on the type of channel -- quantum or classical -- we are investigating. The channel ${\cal C}_{M_C}^\lambda$ consists of the convex combination of the former two channels with the shared memory:
\be\label{memorymix}
{\cal C}_{M_C}^\lambda = \lambda {\cal A}_{M_C}+(1-\lambda){\cal B}_{M_C}.
\ee

In a nutshell, the general form of Parrondo's paradox for communication channels with memory works as follows. Upon sufficiently large number $N$ of individual uses of each of the channels $\am$ and $\bm$, the state of the memory registers $M_A^N$ and $M_B^N$ respectively has the property
\begin{align}
\begin{cases}
\mathbb{E}[M_A^N]<M_A^0\\
\mathbb{E}[M_B^N]<M_B^0,
\end{cases}
\end{align}
where $M_i^0$ for $i=A,B$ denote the initial state of the memory. However, for ${\cal C}_{M_C}^\lambda$ one has
\be
\mathbb{E}[M_C^N]>M_C^0.
\ee
We will further omit the subscript of the memory register $M$ when it does not cause confusion regarding the channel it belongs to. It is rather counterintuitive to expect that the expected value of the memory of the joint channel can increase, if the memory of individual channel decreases with the number of its applications. Previously, this effect has been demonstrated to occur in a variety of physical systems~\cite{harmer_game_1999}, but it has never been considered in the context of information theory.

The gist of the Parrondo's paradox for communication channels consists of constructing channels which take advantage of this unexpected reversal in the  growth of the value of the memory register.

\section{Superactivation of the Quantum and Private Capacity}\label{quantcap}

We now turn to the explicit construction of quantum channels with memory, which demonstrates the superactivation-like effect with inputs independent across the channel while utilizing only erasure channels. 
Each of the channels is the erasure channel ${\cal N}_p$ defined as:
\be
\er {p} (\rho) = (1-p)\rho+p|e\ra\la e|.
\ee
This channel is arguably one of the simplest nontrivial quantum channels. It is known that for $p\in [\half,1]$ the channel has zero quantum and private capacity~\cite{bennett_capacities_1997} and it has strictly positive capacity for all values of $p\in [0,\half)$.

We consider two channels with memory ${\cal A}_M$ and ${{\cal B}_M}$, where $M$ denotes a classical memory subsystem. We show below that each of the channels has zero capacity, but their convex combination results in a channel with positive capacity.

The first channel, ${\cal A}_M$, operates as follows:
\begin{enumerate}
\item $\er {p_a}$ with $p_a = 0.5$ acts on the input state $\rho$.
\item If erasure takes place, the value of $M$ becomes $\wt M = M-1$. Otherwise, $\wt M = M+1$.
\end{enumerate}
The second one, ${\cal B}_M={\cal T}^{M_0}_M\circ \pc_M$, is the composition of two channels $\pc_M$ and ${\cal T}^{M_0}_M$ acting sequentially. The channel $\pc_M$  acts on the input state $\rho$ with one of the erasure channels $\er {p_b}$ or $\er {p_c}$ depending on the state of the classical memory $M$:
\begin{enumerate}
\item Depending on the value of $M$, $\pc_M$ acts as follows:
\be
\pc_M(\rho)=\begin{cases}
\er {p_b}(\rho), \mbox{    if } M\mbox{ mod }3=0;\\
\er {p_c}(\rho), \mbox{    otherwise}.
\end{cases}
\ee
\item If the application of $\er {p_b}$ or $\er {p_c}$ results in erasure then the value of $M$ becomes $\wt M = M-1$. Otherwise $\wt M = M+1$.
\end{enumerate}
The channel, ${\cal T}^{M_0}_M$ acts as follows:
\be
{\cal T}^{M_0}_M(\rho)=\begin{cases}
\er {0}(\rho), \mbox{    if } M>M_0;\\
\er {1}(\rho), \mbox{    if}  M\le M_0,
\end{cases}
\ee
where ${\cal N}_0(\rho)=\rho$, ${\cal N}_1(\rho)=|e\ra\la e|$. Then, ${\cal B}_M = {\cal P}_M\circ {\cal T}_M^{M_0}$, and acts on the input state $\rho$ as follows:
\begin{enumerate}
\item $\pc_M$ acts on the input state $\rho$.
\item Its output is forwarded to ${\cal T}^{M_0}_M$.
\item ${\cal T}^{M_0}_M$ reads out the value of $M$ and if $M\ge M_0$, then $\er {0}$ acts the state. Otherwise, the $\er {1}$ acts. 
\item The output of ${\cal T}^{M_0}_M$ is forwarded to the receiver.
\end{enumerate}

We note that ${{\cal B}_M}$ is a Markov-type channel in the sense that its action depends only on what happened in the previous step. It is also known to be a {\it forgetful channel}. Forgetful channels received a significant amount of attention in the last years due to their tractable properties and simple yet powerful structure: an arbitrary quantum channel with memory can be approximated by a sequence of forgetful channels~\cite{kretschmann_quantum_2005}. Being one of the simplest models of the channels with memory, they admit explicit bounds for their classical and quantum capacity~\cite{datta_coding_2007, bowen_bounds_2005}. In addition to being forgetful it is easy to see that they are also {\it indecomposable} channels~\cite{bowen_bounds_2005}: their operation does not depend on the input state, but only on the internal state of the memory. The following Lemma shows that for certain values of $p_b, p_c$ the channel ${\cal B}_M$ has zero quantum capacity.
\begin{lemma}\label{lemma:bcapacity}
Consider the channel $\pc_M$, which uses erasure channels $\er {p_b},\er {p_c}$ with erasure probabilities $p_b = 0.9+\epsilon$, $p_c=0.25+\epsilon$, $\epsilon=0.01$. The classical memory register $M$ is initialized with $0$. Then the channel ${\cal B}_M$ whose action is defined in steps 1-5 above has 
\be
{\cal Q}({{\cal B}_M})=0.
\ee
\end{lemma}
{\bf Proof:}
as we will show below the {\it effect} of the ${\cal B}_M$ on the input will be that of an erasure channel with extremely high erasure probability, and for each block of output states of finite length which arrive at Bob, Eve can degrade her output to match his. This amounts to establishing that the channel ${\cal B}_M$ is anti-degradable and thus has zero capacity.

It will suffice to establish the statement of the Lemma for the tensor power inputs $\rho^{\otimes n}_{AA^{'}}$ to the channel, because, as it will become evident at the end,neither entangled or classically correlated  inputs across the uses of the channel will not provide any advantage.

The state of the memory $M$ changes only upon the action of the Markov channel ${\cal P}_M$, and despite of the unbounded range of integer values that $M$ can take, its state may be concisely represented by three principal states $\{|i\ra\la i|\}_{i=0}^{2}$, which correspond to those values of $M$, for which $M\mbox{ mod }3=i$. Because our channel is Markov, after a large number $n$ of uses of ${\cal P}_M$ the state of the memory will approach its stationary state $M_{st} = \pi_0|0\ra\la0|+\pi_1|1\ra\la1|+\pi_2|2\ra\la2|$, where $\{\pi_i\}_{i=0}^{2}$ are the probabilities to be in the state $|i\ra\la i|$. From Markov property of the channel it also follows that 
\be\label{statmemdist}
\norm{M- M_{st}}_1\le \delta(n),
\ee 
where $\delta(n)$ vanishes exponentially quickly in the limit $n\to\infty$.

By choosing $M_0$ large enough we ensure that the state of the underlying Markov chain of ${\cal P}_M$ is arbitrary close to the stationary state before any state will get transmitted to Bob via ${\cal T}_M^{M_0}$. Therefore, we will limit the analysis of ${\cal B}_M$ to this case.

We find $\{\pi_i\}_{i=0}^{2}$ from the system of equations for the stationary distribution of the Markov chain:
\begin{align}
\begin{cases}\label{eqa}
(1-p_c)\pi_2+p_c\pi_1&=\pi_0\\
(1-p_b)\pi_0+p_c\pi_2&=\pi_1\\
(1-p_c)\pi_1+p_b\pi_0&=\pi_2\\
\pi_0+\pi_1+\pi_2&=1.
\end{cases}
\end{align}
This gives $\pi_0\approx0.3844$.
The success probability of Alice transmitting the state $\rho$ through ${\cal P}_M$ when the memory is in the stationary state is:
\begin{align*}
Pr(x=\rho)&=(1-p_b)\pi_0+(1-p_c)(\pi_1+\pi_2)\\
&=(1-p_b)\pi_0+(1-p_c)(1-\pi_0)\\
&\approx0.49914,
\end{align*}
where $x$ denotes the state that was output by ${\cal P}_M$.
Now, consider the channel ${\cal B}_M$ in its entirety. In order for any state from ${\cal P}_M$ to be communicated to Bob, the value of $M$ must exceed $M_0$. To understand the behaviour of $M$ we consider a random walk on $\mathbb{Z}$ starting at $0$, and moving left (decreasing the value of $M$ by $1$) with probability $p$, and right (increasing the value of $M$ by $1$) with probability $q=1-p$, where
\begin{align*}
&p=Pr(x=\rho)=0.49914\\
&q=Pr(x=|e\ra\la e|)=0.50086.
\end{align*}
The value of $M$ after $n$ steps is described as the random walk $S_n=\sum_{i=1}^{n}X_i$ of $n$ random variables $\{X_i\}_{i=1}^n$, each of which takes values $+1$ with probability $p$ and $-1$ with probability q.
Note that $\mathbb{E}\li[S_n\pr] = n\mathbb{E}\li[X_1\pr]=-2\alpha n$, where $\alpha = 0.009$, and it linearly decreases with each subsequent transmission.

We are now ready to show that effect of ${\cal B}_M$ on the input state is that of an erasure channel $\er s$, with the probability of erasure $s \gg \half$. Consider the probability of Bob getting a block $L^r_k=l_1...l_k$ of fixed length $k$ through ${\cal B}_M$ on steps $n,n+1,...,n+k-1$ of transmission. Each $l_i$ is either the erasure flag or a faithfully transmitted state. More formally, let $l_{i_1}=...=l_{i_r}=\rho$, with $i_1\le...\le i_r$, $r\ge1$, and $l_j=|e\ra\la e|$ for $j\in\{1,...,k\}\setminus\{i_1,...,i_r\}$, $r\le k$. Because ${\cal B}_M$ is the composition of the two channels ${\cal T}_M^{M_0}$ and ${\cal P}_M$, the probability of transmitting a sequence of states $L_k$ via ${\cal B}_M$ factorizes as follows:
\begin{align*}
Pr(L_k \mbox{ is sent to Bob})=&Pr({\cal P}_M\mbox{ outputs }L_k)\\&\times Pr(S_{i_1}\ge M_0,...,S_{i_r}\ge M_0).
\end{align*}
We can upper-bound the right hand side above as:
\begin{align*}
Pr(\mbox{Bob receives } L_k)&\le Pr(S_{i_1}\ge M_0,...,S_{i_r}\ge M_0)\\&
\le Pr(S_{n}\ge M_0).
\end{align*}
To estimate $Pr(S_{n}\ge M_0)$ we make use of the Hoeffding's inequality~\cite{hoeffding_probability_1963}:
\begin{align}
Pr(S_{n}\ge M_0) &= Pr(S_{n}-\mathbb{E}\li[S_n\pr]\ge M_0-\mathbb{E}\li[S_n\pr])
\\&= Pr(S_{n}+2\alpha n\ge M_0+2\alpha n)\\
&\le \exp\li({-\frac{2(M_0+2\alpha n)^2}{4n}}\pr).
\end{align}
Therefore,
\be 
Pr(\mbox{Bob receives } L_k)\le\exp\li(-2\alpha^2 n\pr).
\ee
To ensure that entangled inputs do not help, it is enough to show that the effect of ${\cal B}_M$ on the input is that of a memoryless erasure channel. Instead of the fixed block $L_k$ we consider the general form of the block trainsmitted to Bob of length $k$, $S_k^B=s_1...s_k$ and show that for each $i$ the port $s_i$ was obtained by some erasure channel with the probability of erasure greater than $0.5$. The $i$-th position in $S_k^B$ contains the mixture $s_i^B=r\rho + (1-r)|e\ra\la e|$, where from the above analysis we get $r=\exp{(-(n-k+i)C)}$ for some constant $C>0$, which depends only on $p_b$ and $p_c$.
Picking large enough $M_0$ we can guarantee $r\ll 0.5$ for all $k$, and the action of ${\cal B}_M$ on the input state is that of $\er {1-r}$ with $r\to 0$ when $n\to\infty$. 

One can understand this as follows: with each step, the expectation of the random walk $\mathbb{E}\li[S_n\pr]$ travels towards $-\infty$ with the speed which is linear in $n$, i.e. the difference $|M_0-S_n|$ grows linearly. Central limit theorem states that after $n$ steps $S_n\in(-\sqrt{n}+\mathbb{E}\li[S_n\pr];\mathbb{E}\li[S_n\pr]+\sqrt{n})$ with high probability, but this is not enough, because we need to guarantee that deviating from the average linearly is highly unlikely. Large deviation bound confirms this intuition: the probability of being polynomially far from this region after $n$ steps vanishes exponentially.

Having established that regardless of input the total effect of ${\cal B}_M$ on each of the ports of $S_k^B$ is tantamount to that of a particular erasure channel with probability of erasure $\displaystyle (1-r)\to1$, the optimal coding for ${\cal B}_M$ must be the same as the optimal coding for the erasure channel $\er s$. We know that its capacity is achieved on tensor product inputs~\cite{bennett_capacities_1997}.

Lastly, what remains to be proved is that ${\cal B}_M$ has zero capacity. It suffices to show that the complementary channel ${\cal B}_M^c$ is degradable (definition of degradability for quantum channels with memory and its implications is given in Supplementary Section), which implies that ${\cal B}_M$ is anti-degradable and thus has zero quantum capacity. This amounts to showing that for each output block $S_k^B$ of length $k$ at Bob's possession there exists degrading map ${\cal D}_k$ which can degrade Eve's output $S_k^E$ to match Bob's. Again, we will consider the stationary case (number of uses of the channel $n\gg1$), because in the non-stationary case the probabilities of getting the state or the erasure flag will be weighted by $\delta(n)$ from~\eqref{statmemdist}, which can be made arbitrarily small. Also, we consider the non-trivial regime when $M_0>M$ and the channel ${\cal T}_M^{M_0}$ becomes ${\cal N}_0$.
The $i$-th position of $S_k^E$ contains the mixture $s_i^E=(1-r)\rho + r|e\ra\la e|$. Showing that Eve can degrade every position of the $S_k^E$ to match $S_k^B$ is sufficient to establish degradability of the whole block. Let 
${\cal D}_{k,i}$ be the degrading map that acts on the $i$-th position of the block. Then, it can be seen as performing the coin flip and with probability $t$ returning $s_i^E$ and with probability $1-t$ returning $|e\ra\la e|$, i.e. performing the map $(1-r)\rho + r|e\ra\la e|\to t((1-r)\rho + r|e\ra\la e|)+(1-t)|e\ra\la e|= t(1-r)\rho + (1-t(1-r))|e\ra\la e|$. This yields $t=r/(1-r)$ and the degrading map for $S_k^E$ has the form ${\cal D}_k = {\cal D}_{k,1}\circ...\circ {\cal D}_{k,k}$.
\qed

Remarkably, transmitting quantum states using the convex mixture of two zero-capacity channels ${\cal A}$ and ${\cal B}_M$ in which both of the channels have access to the shared memory register $M$ enables the sender to convey quantum information reliably. 

More precisely, consider the channel ${\cal C}_M^\lambda = \lambda {\cal A}_M + (1-\lambda){{\cal B}_M}$, with $\lambda\in(0,1)$ and the classical memory $M$ shared between the channels. The Lemma below shows that the quantum capacity of such convex mixture for $\lambda = 0.5$ is strictly positive -- the capacity of ${\cal C}_M^\lambda$ experiences superactivation-like effect. Note that this is qualitatively different from establishing the non-convexity condition for the quantum capacity which states that the capacity of the channels in the mixture is larger than the mixture of individual capacities of the respective channels. Non-convexity is merely the necessary condition for superactivation, which presents the strongest violation of additivity, because the channels in the mixture need not have zero capacity.

We now show that for $\lambda = 0.5$ such convex mixture of zero-capacity erasure channels with memory has positive capacity.
\begin{lemma}\label{lemma:finale}
Consider the channel ${\cal C}_M^{0.5} = 0.5 {\cal A}_M + 0.5{{\cal B}_M}$, where the underlying erasure channels of ${\cal B}_M$ (which form ${\cal P}_M$) are $\er {p_b},\er {p_c}$ with erasure probabilities $p_b = 0.9+\epsilon$, $p_c=0.25+\epsilon$, $\epsilon=0.01$ and $M_0$ is large. The state of the memory register $M$ is initially $0$. Then,
\be
{\cal Q}\li({\cal C}_M^{0.5}\pr)>0.
\ee
\end{lemma}
{\bf Proof:} it is sufficient to restrict ourselves to the tensor product inputs and the stationary state of the memory $M_{st}$ by picking a large enough $M_0$ as in the proof of Lemma~\ref{lemma:bcapacity}. The value of $M$ after $n$ steps of transmission will be determined by the random walk $S_n = \sum_{i=1}^{n}X_i$ where each of the random variables $\{X_i\}_{i=1}^n$ takes values $+1$ with probability $p$ and $-1$ with probability $q=1-p$ with $p$ and $q$ computed below. Unlike the case with ${\cal B}_M$, the memory register of ${\cal C}_M^{0.5}$ can change in two distinct ways: when $M \mbox{ mod 3}=0$ then half of the time ${\cal P}_M$ has acted on the input with $\er {p_b}$ transmitting the input perfectly with probabilities $1-p_b$, and half of the time with $\er {p_a}$, which sent the state intact with probability $1-p_a$. Similarly, in other cases, half of the time $\er {p_c}$ acted on the input state transmitting the input perfectly with probability $1-p_c$, or half of the time $\er {p_a}$ sent the state perfectly with probability $1-p_a$. This results in new probabilities $r_b$, $r_c$ of successfully sending the state  when the memory is divisible by $3$ and when it is not, respectively:
\begin{align}
r_b = \half (1-p_a)+\half(1-p_b),\\
r_c = \half (1- p_a)+\half(1- p_c).
\end{align}
Then solving the system~\eqref{eqa} with new probabilities of success $(1-p_b)=r_b$ and $(1-p_c)=r_c$ we find $\wt\pi_0=0.345$ and

\begin{align*}
&p=Pr(x_k=\rho)=r_b\wt\pi_0+r_c(1-\wt\pi_0)\approx 0.5078,\\
&q=1-p\approx 0.4922.
\end{align*}
Then $\mathbb{E}\li[S_n\pr] = n\mathbb{E}\li[X_1\pr]=2\alpha n$, where $\alpha = 0.0078$. Unlike the random walk, which underpins ${\cal B}_M$ in Lemma~\ref{lemma:bcapacity}, this random walk is biased in the positive direction, and for a sufficiently large number $n$ of uses of ${\cal C}_M^{0.5}$ the value of $M$ will be larger than any fixed constant $M_0$.  To formalize this intuition, define $x_i = \rho$, $i=1,...n$ to be Alice's input to ${\cal C}_M^{0.5}$, and $y_i$, $i=1,...,n$ to be Bob's output. The probability of Alice to transmit the input state $\rho$ to Bob successfully on the $n$-th use of the channel is given by:
\begin{align}
Pr(y_n&=\rho) \\
&= Pr(x_n=\rho)(1-Pr(S_n\le M_0)) \\
&=p(1-Pr(\widetilde S_n\ge -M_0))\\
&=p(1-Pr(\widetilde S_n-\mathbb{E}[\widetilde S_n]\ge -M_0-\mathbb{E}[\widetilde S_n]))\\
&=p(1-Pr(\widetilde S_n+2\beta n\ge -M_0+2\beta n))\\
&=p\li(1-\exp{\li(-\frac{(-M_0+2\beta n)^2}{4n} \pr)}\pr)\label{hoeffd},
\end{align}
where $\widetilde S_n=\sum_{i=1}^{n}\widetilde X_i$ is the symmetric reflection of the random walk $S_n$ with $X_i$ being $+1$ with probability $q$ and $-1$ with probability p and $\mathbb{E}\widetilde S_n=2\beta n$, $\beta = -\alpha$. We applied Hoeffding's inequality in Eqn.~\eqref{hoeffd}.

After some finite number of steps $n_0$ the channel ${\cal T}_M^{M_0}$ starts acting as the identity channel, and the probability of sending the state $\rho$ to Bob is given by $Pr(y_n=\rho)$. Note that $\lim_{n\to\infty}Pr(y_n=\rho)=\lim_{n\to\infty}p\li(1-\exp{\li(-\frac{(-M_0+2\beta n)^2}{4n} \pr)}\pr)=p$. For large (finite) $n$ the effect of ${\cal C}_M^{0.5}$ is that of an erasure channel $\er {s}$ with $s<0.5$, which has positive capacity.
\qed
The simple structure of the underlying channels in the construction of ${\cal A}$, ${\cal B}_M$ and ${\cal C}_M^{0.5}$ makes it possible to make a similar statement about their private capacity. From the proofs of Lemma 1 and 2 it immediately follows that the private capacity of the two channels and their convex combination can be superactivated:
\begin{cor}
For the quantum channels $\cal A$, ${\cal B}_M$ and ${\cal C}_M^{0.5}$ we have:
\be
\begin{cases}
{\cal P}({\cal A})=0\\
{\cal P}({\cal B}_M)=0\\
{\cal P}({\cal C}_M^{0.5})>0.
\end{cases}
\ee
\end{cor}

\section{Superactivation of the Capacity of Classical Channel with Memory}\label{classcap}

The purpose of this section is to show that using Parrondo's paradox one can construct classical channels with memory for which one can demonstrate superactivation of the capacity. 

Consider $\am(x)$ to be the natural generalization of the binary symmetric channel with memory and probability $p$ of confusing the input $x\in \{0,1\}$:
\be
\begin{cases}
\mbox{$\am (0)=0$ with $p=0.5$ and $\wt M_A = M_A+1$}\\
\mbox{$\am (0)=1$ with $p=0.5$ and $\wt M_A = M_A-1$}\\
\mbox{$\am (1)=1$ with $p=0.5$ and $\wt M_A = M_A+1$}\\
\mbox{$\am (1)=0$ with $p=0.5$ and $\wt M_A = M_A-1$}
\end{cases}
\ee
The capacity of $\am$ is 
\be
{\cal C}(\am)=1-H(0.5)=0,
\ee
where $H(p)$ is the binary entropy function.

The second channel, $\bm$, has identical construction to the one described in Section~\ref{quantcap} with the only difference that instead of quantum states the channel takes input $x\in\{0,1\}$. Following the proof of Lemma~\ref{lemma:bcapacity} up to the step where we prove degradability amounts to establishing that the probability of successful decoding for the rate $R>0$ decays exponentially. This implies that $\bm$ has vanishing capacity and $\lim_{n\to\infty}{\cal C}({\cal B}_{M_B}^{(n)})=0$, where the superscript indicated the number of uses of $\bm$. Now, consider the channel $\cm^{0.5}$, comprised of the channels ${\cal A}_{M_C}$ and ${\cal B}_{M_C}$ as in Eqn.~\eqref{memorymix}. After sufficiently many uses of the channel, the memory of the channel $\cm^\lambda$ becomes $M_C\gg M_0$ and we are left with the mixture of
\be
\wt{\cal C}^{0.5}_{M_C} = 0.5{\cal A}_{M_C} + 0.5{\cal P}_{M_C}.
\ee

Following the calculation of Lemma~\ref{lemma:finale}, it is straightforward to see that ${\cal C}({\cal C}^{0.5}_{M_C} )>0$.

\section{Discussion}

In our paper we have shown how Parrondo's paradox naturally leads to constructions, which demonstrate the superactivation of the capacity for classical and quantum channels with memory. We constructed two quantum zero-capacity channels which consist of the combination of the simplest erasure channels. Adding classical memory to the construction made it possible to superactivate the capacity of the mixture of those channels. This is especially surprising in the light of the fact that the quantum capacity of erasure channels cannot be superactivated by one or more memoryless erasure channels with zero capacity. Adding classical memory lifts this necessary requirement, which manifests itself for the known examples of superactivation in the memoryless case.

We have also shown the analogous construction for classical communication channels with memory demonstrating superactivation of their classical capacity. The effect which was only known to exist for quantum capacity of the memoryless quantum channel has now become possible for other types of capacity of quantum as well as classical communication channels once we allowed classical memory.

There were two distinct kinds of erasure channels $\er {p}$ used in the construction of the quantum channels with memory: degradable ($p<0.5$) and anti-degradable ($p\ge 0.5$). Despite similar description, their properties are significantly different. The former class of channels does not have a tractable analytical characterization. The chanels of this kind do not even form a closed set. On the other hand, the class of all anti-degradable channels, being closed and convex, admits tractable characterization. For many practical applications one may send quantum information through the channel which may be either degradable or anti-degradable without full knowledge of which of the two classes it belongs to. The natural example is sending quantum information through the erasure channel when there is some uncertainty about its erasure probability or when erasure probability varies. The open question is to find the capacity when the transmission is performed using channels which belong to one of the classes above.

In the context of memoryless quantum channels, it is known that the quantum capacity can be superactivated. It is plausible that the private capacity can be superactivated as well, but this question is still open despite attempts to resolve it~\cite{smith_extensive_2009, li_private_2009}. 
Finding a suitable way of approximating the channels with memory in our construction with their memoryless counterparts will lead to resolution of this question.

The nontrivial channel in our construction, $\bm$, consists of the composition of two channels, ${\cal P}_M$ and ${\cal T}_M^{M_0}$. However, the use of ${\cal T}_M^{M_0}$ is rather artificial. It was necessitated by the fact that it was not possible to show that the channel ${\cal P}_M$ alone has zero capacity. It makes the construction insensitive to the nature of the underlying channels -- quantum or classical -- used for superactivation. Therefore, an open question is whether there exists a way to achieve the genuine quantum superactivation effect for quantum channels with memory which is impossible for the classical channels with memory.

{\it Acknowledgments.} S.S. thanks Jonathan Oppenheim for his comments on the early draft of this paper and Toby Cubitt for clarifying several aspects of degradable channels. S.S. thanks Centre for Quantum Information and Foundations for its support during his Ph.D. studies. 

\section{Supplementary Section}
\subsection{Degradable channels with memory}
The notion of degradable channel, first introduced by~\cite{devetak_capacity_2005} was only considered in the context of memoryless channels. Here we provide a natural generalization of the notion of degradability to the channels with memory. For the latter, one cannot talk about the Choi-Jamio\l kowski state of the channel. Therefore, a natural generalization of the notion of degradability (and anti-degradability) makes use of the output of the channel with memory after the finite number of its uses.
\begin{defn}
The channel $\chn_M$ with memory register $M$ is degradable if 
\be
\forall n \exists \mbox{ CPTP map }{\cal D}^{(n)}: {\cal N}_M^{c,(n)} = {\cal D}^{(n)}\circ {\cal N}_M^{(n)},
\ee
\end{defn}where ${\cal N}_M^{(n)}$ and  ${\cal N}_M^{c,(n)}$ represent $n$ uses of $\chn_M$ and its complementary channel respectively. Similarly, one defines anti-degradable channel with memory:
\begin{defn}
The channel $\chn_M$ with memory $M$ is anti-degradable if 
\be
\forall n \exists\mbox{ CPTP map } {\cal D}^{(n)}: {\cal N}_M^{(n)} = {\cal D}^{(n)}\circ {\cal N}_M^{c,(n)}
\ee
\end{defn}

Analogously to the memoryless case, anti-degradable channel with memory has zero quantum capacity:
\begin{obs}
Let $\chn_M$ be the anti-degradable channel with memory and ${\cal N}_M^{(n)}$ denotes its use $n$ times. Then for all inputs $\sigma$ the following is true:
\begin{enumerate}
\item $I_c(\sigma,{\cal N}_M^{(n)})\le0$.
\item ${\cal Q}({\cal N}_M^{(n)})=0$.
\item ${\cal Q}({\cal N}_M)=0$.
\end{enumerate}
\end{obs}
The proof of the above observations is identical to the proof of Proposition 1 in~\cite{holevo_entanglement-breaking_2008}.

\bibliographystyle{abbrv}

\end{document}